\documentclass[%
 reprint,
 amsmath,amssymb,
 aps,
 pre
]{revtex4-2}

\usepackage{graphicx} 
\usepackage{dcolumn}  
\usepackage{bm}       
\usepackage{mathtools} 
\usepackage{color}
\usepackage[normalem]{ulem} 
\definecolor{editgray}{rgb}{0.55,0.55,0.55}
\definecolor{editpurple}{rgb}{0.50,0.00,0.60}
\definecolor{editgreen}{rgb}{0.00,0.55,0.25}
\begin{document}

\title{An Investigation of the Channel Capacity of Bacterial Chemotactic Sensors \\ for Low Chemoattractant Concentrations}

\author{Ziyi Cui}
\affiliation{Pioneer Academics, Philadelphia, PA, United States of America}

\author{Sarah Marzen}
\email{smarzen@natsci.claremont.edu}
\affiliation{Department of Natural Sciences, Pitzer and Scripps College, Claremont, CA, USA}
\affiliation{Kravis Department of Integrated Science, Claremont McKenna College, Claremont, CA, USA}
\affiliation{National Institute for Theory and Mathematics in Biology, Chicago, IL, USA}

\date{\today}

\begin{abstract}
Bacterial chemotactic sensing converts noisy chemical signals into running and tumbling. We analyze the static sensing limits of mixed Tar/Tsr chemoreceptor clusters in individual \textit{Escherichia coli} cells using a heterogeneous Monod--Wyman--Changeux (MWC) model. Across a seven-dimensional parameter sweep, we compute three sensing-performance metrics -- channel capacity, dynamic range, and effective Hill coefficient -- in the limit that the cells are constantly in such low concentrations of chemoattractant that they need not adapt to new baseline chemoattractant concentration levels. What results are an upper bound on the trajectory mutual information rate given by the static capacity divided by a finite perceptual timescale, a quantitative understanding of the tight connection between channel capacity and dynamic range, and the finding that in this regime channel capacity is well described by a closed-form ceiling depending only on the receptor's baseline activity, which every wild-type and mutant strain in our sample achieves to within a few percent. In more realistic scenarios, adaptation plays a larger role and so do the exact temporal dynamics of chemoattractant concentrations seen by bacteria as they swim. This manuscript thus points to the importance of mapping out naturalistic chemoattractant concentration statistics in the wild as has been done for natural scene statistics.

\end{abstract}



\maketitle

\section{Introduction}

Sensing is fundamental to life. Across all domains of biology, sensing is a noisy conversation between the environment and an organism's decision-making. With bad sensors can come bad decisions, and so evolution might have driven organisms to maximize sensors' information processing capabilities for survival.

A long-standing biophysical hypothesis thus theorizes that evolution pushes sensory pathways to maximize information transfer from environment to sensor output. For instance, measurements indicate that fruit fly embryos transmit information on gene regulation from chemical gradients at approximately $90\%$ of the theoretical maximum \cite{tkavcik2008information}. Nicotinic acetylcholine receptors in the uncoupled limit appear to have maximized their channel capacity \cite{marzen2013statistical}, though see Ref.~\cite{goldfein2026predictions}; for Monod-Wyman-Changeux molecules, high channel capacity is correlated with maximal cooperativity and maximal dynamic range \cite{martins2011trade}. However, potassium voltage-gated channels within animals have been found to operate far from an information-maximizing regime and have evolved toward lower information transfer \cite{duran2023not}. Similarly, single mammalian cells only transmit sufficient information for binary decision-making, and additional information is limited by an upstream bottleneck \cite{cheong2011information}. Finally, evolutionary channels can only transmit a few bits of information \cite{soriano2023well}.

A model system for studying this hypothesis is bacterial chemotaxis. The chemotactic receptors, responsible for the movement of bacterial cells in response to environmental chemical stimuli, have been the subject of a number of beautiful models \cite{marzen2013statistical,tu2013quantitative,mello2003quantitative,mello2005allosteric}. In order to survive, \textit{E.\ coli} cells must take in information about chemoattractant concentration \cite{berg1977physics} so as to move to maximize chemoattractant intake.
Recent measurements in \textit{E.\ coli} cells have demonstrated that they operate close to optimal information transfer for gradient-climbing \cite{mattingly2021escherichia}. A recent study showed that bacteria optimize predictive information transfer given resource constraints \cite{tjalma2023trade}. Furthermore, basic theoretical analyses say that populations of bacteria should maximize predictive information subject to constraints on resources \cite{marzen2018optimized}. 

However, no studies have addressed the channel capacities of individual bacteria. While some studies on \textit{E.\ coli} chemotactic receptors focus on the population level and dynamical behavior of cells \cite{martins2011trade,marzen2018optimized}, the present study provides an analysis of the chemotactic ability of a single \textit{E.\ coli} cell. Specifically, the mixed Tar/Tsr chemoreceptor clusters within the \textit{E.\ coli} cell are modeled with a heterogeneous Monod-Wyman-Changeux molecule \cite{marzen2013statistical} under the assumption of adaptation to negligible background chemoattractant concentration \cite{celani2010bacterial}. Three sensor performance metrics (channel capacity, dynamic range, and effective Hill coefficient) are obtained across a seven-dimensional parameter sweep, and their values for wild-type and mutant strains are compared to the maximum possible values over the possible parameter values that could have been selected by evolution. The results point to the potential importance of accounting for adaptation and naturalistic chemoattractant concentration statistics when calculating information transfer, though here we compute a static channel capacity, an upper bound on information per measurement in a low-background regime, which can be converted into a bound on information rate only after accounting for the perceptual timescale, which we take to be the pathway response time (Sec.~\ref{sec:static}).

\section{Background}

\subsection{Bacterial Chemotaxis}

\begin{figure}[h!]
    \centering
    \includegraphics[width=\linewidth]{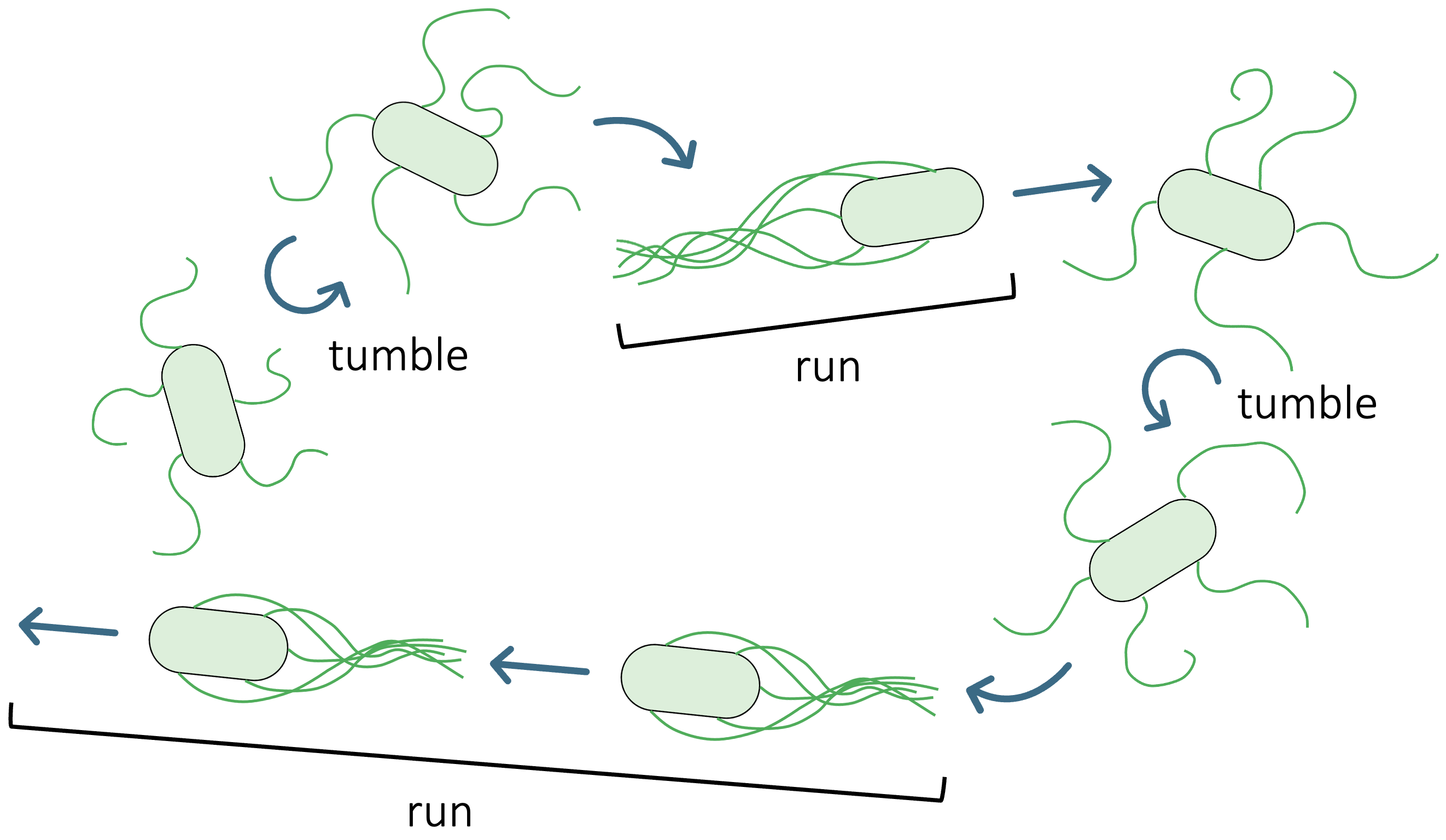}
    \caption{Schematic demonstrating the run--tumble behavior of \textit{E.\ coli}.}
    \label{run and tumble}
\end{figure}

Motile bacteria regulate their motion in response to spatial variations in chemical concentrations so that, on average, trajectories are biased up attractant gradients and down repellent gradients. This bias does not require the cell to know the global gradient. Instead, as the cell swims, it continuously samples the local environment and adjusts its behavior based on changes in concentration it has recently experienced. Bacterial chemotaxis is characterized by an alternating sequence of runs and tumbles \cite{marzen2013statistical}. During a run, several flagella rotate in the same direction, form a bundle, and propel the cell along an approximately straight path. A tumble occurs when the cell reorients itself by breaking the bundle and rotating its flagella in different directions (Fig.~\ref{run and tumble}). 

The chemical signals that bias this run--tumble behavior are called chemoattractants or chemorepellents, which bind to chemoreceptors embedded in the cell membrane. In \textit{E. coli}, the dominant chemoreceptor types are Tar and Tsr. Tar primarily recognizes aspartate and its analogues such as MeAsp, whereas Tsr primarily recognizes serine. At the cell pole, many Tar and Tsr receptors organize with the adaptor CheW and the histidine kinase CheA in a receptor cluster whose members switch conformations collectively. The cluster alternates between an inactive state, in which CheA kinase activity is suppressed, and an active state, in which CheA autophosphorylation from ATP is stimulated \cite{marzen2013statistical}. Mixed Tar/Tsr composition tunes the cluster's sensitivity to distinct ligands, and the cooperative coupling ensures that modest changes in ligand occupancy can produce large changes in the cluster's output.

\begin{figure}[h!]
    \centering
    \includegraphics[width=0.7\linewidth]{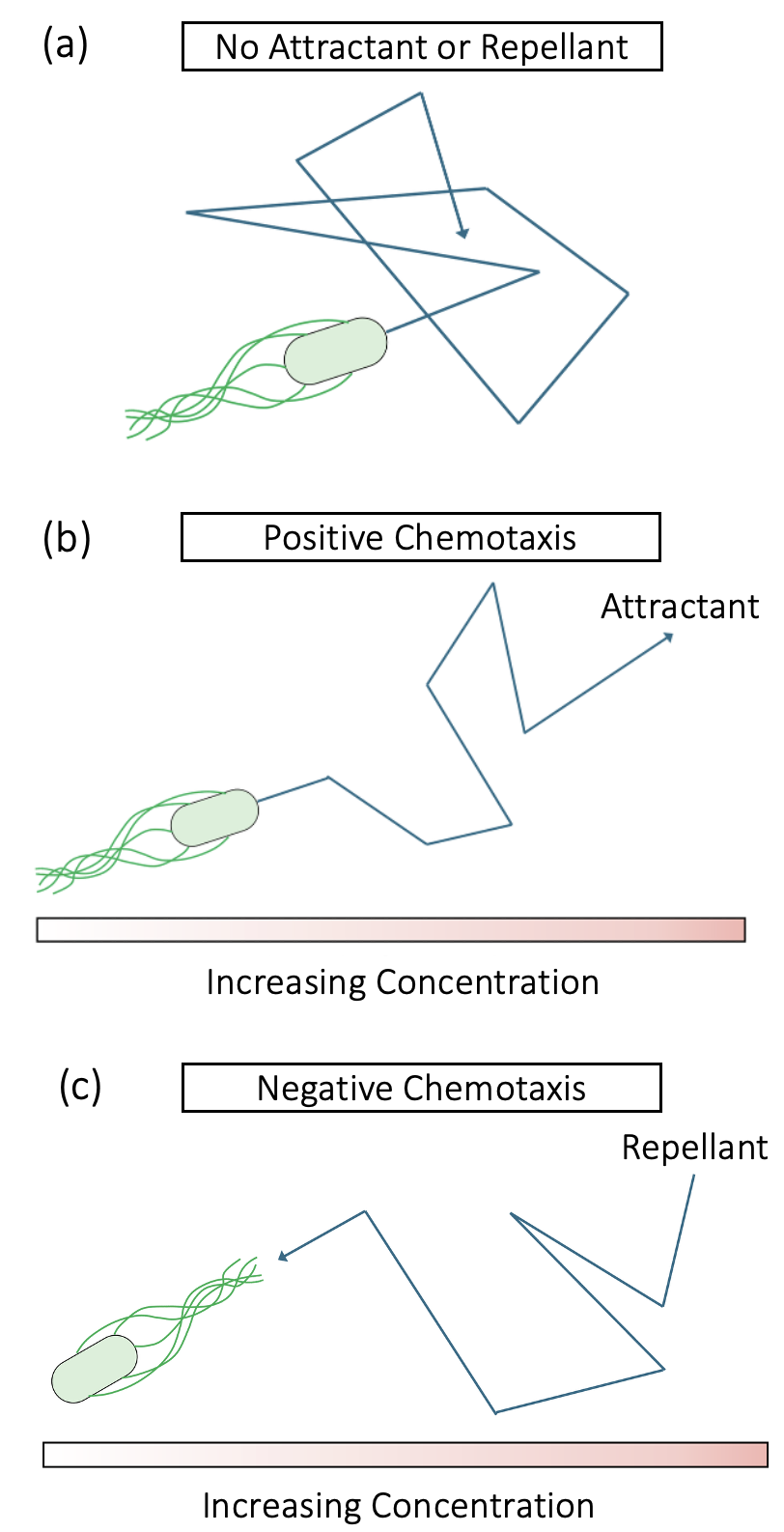}
    \caption{Schematic demonstrating bacterial chemotactic behavior and receptor response. (a) In a spatially uniform environment (no attractant or repellent), the run--tumble trajectory is an unbiased random walk with no net drift. (b) Positive chemotaxis in an attractant gradient: run segments are longer when the cell swims up the gradient and shorter when it swims down, yielding a net displacement toward higher concentration. Bar indicates increasing concentration. (c) Negative chemotaxis in a repellent gradient: more frequent reorientation when moving up-gradient biases motion toward lower concentration.}
    \label{chemotaxis}
\end{figure}

The downstream link to motility is mediated by the CheY response regulator. Phosphorylated CheA transfers its phosphate to CheY, producing CheY-P. CheY-P binds to the flagellar motor and increases its clockwise rotation, causing a tumble. In contrast, when the CheA output is reduced, CheY-P levels drop and the flagellar motor rotates counterclockwise to induce a run. For attractants such as MeAsp or serine, ligand binding stabilizes the inactive cluster state and thereby lowers CheA/CheY-P signaling, biasing the motor toward longer runs; decreases in attractant or the presence of repellents shifts the balance toward the active state and increases the likelihood of reorientation. As a result, bacterial chemotaxis produces net motion toward higher concentrations of attractants and away from higher concentrations of repellents (Fig.~\ref{chemotaxis}).

\subsection{Monod-Wyman-Changeux Model of Bacterial Chemotactic Receptors}
\label{sec:mwc}

Consider a receptor cluster composed of $N_1$ Tar receptors (class $i=1$) and $N_2$ Tsr receptors (class $i=2$). The cluster switches collectively between two conformations $s\in\{I,A\}$ (inactive $I$, active $A$). At an external ligand concentration $c$, binding to receptor class $i$ in conformation $s$ is characterized by the state-dependent dissociation constant $K_{d}^{(s),i}$. At any instant, some integer number of Tar receptors and Tsr receptors are bound to ligand. Let $\varepsilon_s$ denote the conformational free energy of state $s$ and $\beta=1/(k_B T)$.

We define $p(c)$ as the probability that the cluster is in the active conformation at ligand concentration $c$. From Ref. \cite{marzen2013statistical}, we have
\begin{equation}
p(c)=
\frac{
  e^{-\beta\varepsilon_A}\Bigl(1+\tfrac{c}{K_{d}^{(A),1}}\Bigr)^{N_1}
  \Bigl(1+\tfrac{c}{K_{d}^{(A),2}}\Bigr)^{N_2}
}{
  \splitdfrac{%
    e^{-\beta\varepsilon_A}\Bigl(1+\tfrac{c}{K_{d}^{(A),1}}\Bigr)^{N_1}
    \Bigl(1+\tfrac{c}{K_{d}^{(A),2}}\Bigr)^{N_2}
  }{%
    {}\,+\,e^{-\beta\varepsilon_I}\Bigl(1+\tfrac{c}{K_{d}^{(I),1}}\Bigr)^{N_1}
    \Bigl(1+\tfrac{c}{K_{d}^{(I),2}}\Bigr)^{N_2}
  }
}
\label{MWC old}
\end{equation}

\begin{figure}[h!]
    \centering
    \includegraphics[width=1\linewidth]{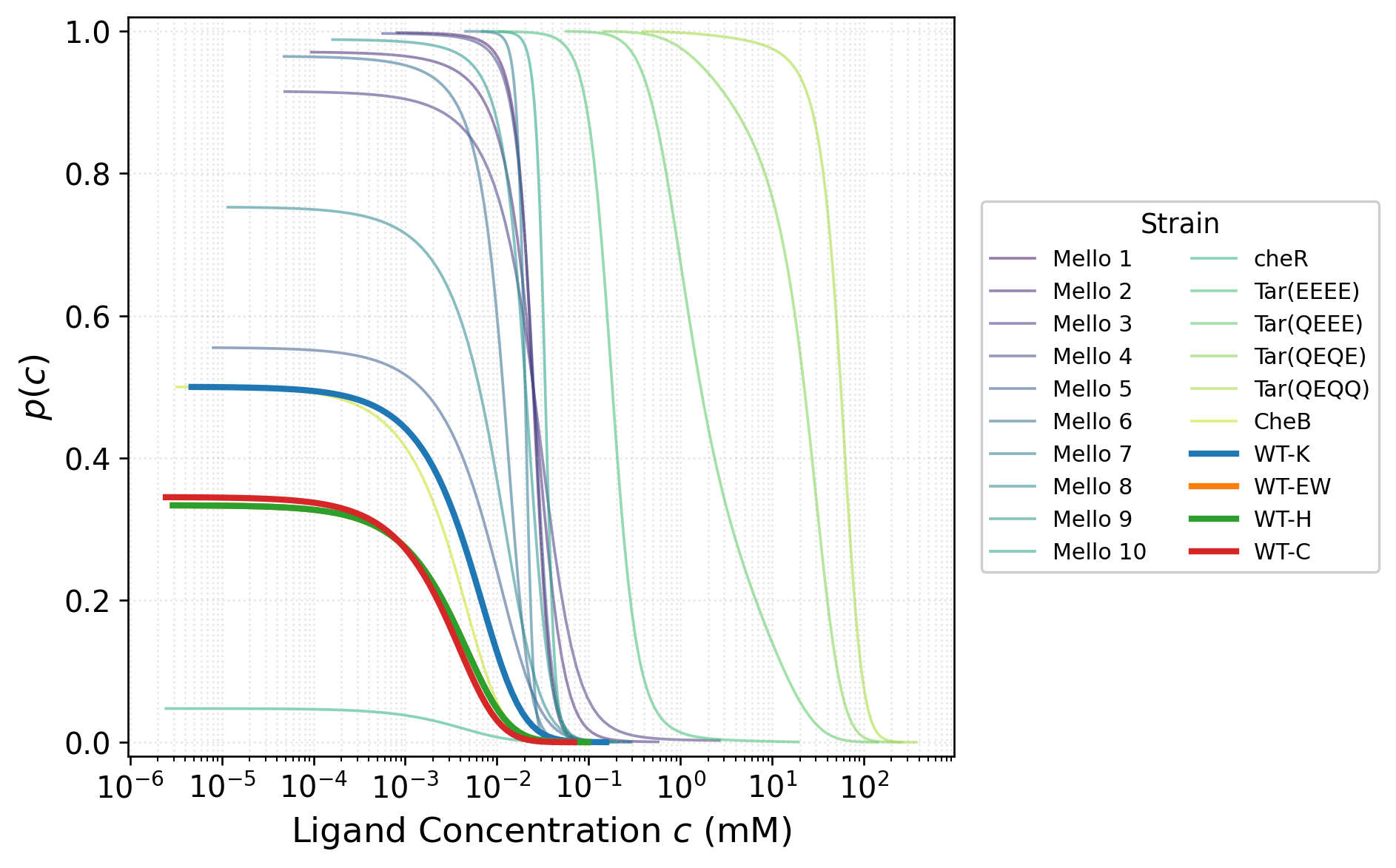}
    \caption{Probability of the active state of a Tar/Tsr receptor cluster for the attractant MeAsp at all twenty strain operating points in Table~\ref{tab:1}. Normalized activity decreases monotonically with MeAsp concentration on a log scale because ligand stabilizes the inactive state. Wild-type strains are shown in saturated colors; lab-grown mutants in lighter shades.}
    \label{MWC curve}
\end{figure}

Define the allosteric constant $L_0$, which represents the ratio of the inactive and active state statistical weights in the absence of ligand, given as
\begin{equation}
L_0 \equiv e^{-\beta(\varepsilon_I-\varepsilon_A)}.
\label{allosteric}
\end{equation}
By substituting Eq.~\ref{allosteric} into Eq.~\ref{MWC old} and dividing the numerator and denominator by
$e^{-\beta\varepsilon_A}\!\left(1+\frac{c}{K_d^{(A),1}}\right)^{N_1}\!\left(1+\frac{c}{K_d^{(A),2}}\right)^{N_2}$ yields
\begin{equation}
p(c)=\frac{1}{1+L_0
\left(\frac{1+c/K_d^{(I),1}}{1+c/K_d^{(A),1}}\right)^{N_1}
\left(\frac{1+c/K_d^{(I),2}}{1+c/K_d^{(A),2}}\right)^{N_2}}.
\label{MWC}
\end{equation}
For attractants with $K_{d}^{(I),i}<K_{d}^{(A),i}$, $p(c)$ decreases monotonically with $c$ (Fig.~\ref{MWC curve} with parameters from Table \ref{tab:1}). Thus, the receptor cluster is more likely to be inactive at attractant concentrations, which allows the \textit{E.\ coli} to have a run-biased trajectory up the attractant gradient.

Adaptation \cite{tu2013quantitative} is ignored not only for simplicity, but also because in natural environments with large bacterial populations, chemoattractants and chemoattractant gradients constantly vanish \cite{celani2010bacterial}, so we focus on a low-background regime. In this paper, ``static'' refers to the instantaneous mutual information (defined in Sec.~\ref{sec:capacity_background}) between ligand concentration and receptor-cluster activity at a fixed receptor modification (methylation) state \cite{tostevin2009mutual,mattingly2021escherichia}; operationally, each strain's $L_0$ value in Table~\ref{tab:1} specifies such a fixed state and sets the baseline activity $p_0 = 1/(1+L_0)$. This description applies on timescales long compared to receptor-cluster equilibration but short compared to the adaptation time, which is roughly $10\ \mathrm{s}$ (measured as $9.9\pm0.3\ \mathrm{s}$ in Ref.~\cite{mattingly2021escherichia}). For a precisely adapting system, the activity returns to the same baseline under any sustained constant concentration, so the infinite-time dose--response curve is flat and carries no information, and any nonzero channel capacity must therefore be defined on timescales shorter than the adaptation time. Under the low-background assumption, we treat the receptor cluster as staying near its zero-background adapted methylation state. In dilute environments, encounters with attractant last on the order of a run ($\sim 1\ \mathrm{s}$), far shorter than the adaptation time, so the methylation state remains near its zero-background adapted level, corresponding to the operating point tabulated for each strain. Slow methylation fluctuations \cite{sartori2011noise,colin2017multiple} can drift the operating point; in the MWC model used here, methylation enters through the allosteric constant $L_0$ and hence the baseline activity $p_0$, so such drift changes which operating point the receptor cluster occupies without changing the capacity ceiling we derive in Sec.~\ref{sec:capacity_results}, although the information actually achieved by a cell whose operating point wanders may be lower. Future work will treat adaptation dynamics explicitly.

\begin{table}[h!]
\centering
\caption{Strain-specific sensor parameters. The four wild-type operating points WT-K, WT-EW, WT-H, and WT-C are reported by Keymer 2006 \cite{keymer2006chemosensing}, Endres \& Wingreen 2006 \cite{endres2006precise}, Hansen 2008 \cite{hansen2008chemotaxis}, and Clausznitzer 2010 \cite{clausznitzer2010chemotactic}, respectively. The sixteen lab-grown receptor-modification mutants comprise Mello \& Tu's ten strains \cite{mello2005allosteric}, five Keymer mutants (cheR and Tar(EEEE/QEEE/QEQE/QEQQ)) \cite{keymer2006chemosensing}, and the Clausznitzer CheB$^{\text{D56E}}$ mutant \cite{clausznitzer2010chemotactic}. The non-integer counts $(N_1, N_2)$ for WT-C and the CheB mutant are expression-weighted averages of the published Tar:Tsr expression ratio $1{:}1.4$ and total cluster size $N = 17.5$ at zero ambient concentration.}
\label{data}
\label{tab:1}
\scriptsize
\setlength{\tabcolsep}{2pt}
\begin{ruledtabular}
\begin{tabular}{lccccccc}
Strain & $N_1$ & $N_2$ & $L_0$ & $K_{d}^{(I),1}$ & $K_{d}^{(A),1}$ & $K_{d}^{(I),2}$ & $K_{d}^{(A),2}$ \\
\hline
WT-K        & 5    & 10    & $1.000$                & 0.02 & 0.5 & 100 & $10^6$ \\
WT-EW       & 6    & 12    & $2.000$                & 0.02 & 0.5 & 100 & $10^6$ \\
WT-H        & 6    & 13    & $2.000$                & 0.02 & 0.5 & 100 & $10^6$ \\
WT-C        & 7.29 & 10.21 & $1.900$                & 0.02 & 0.5 & 100 & $10^6$ \\
\hline
Mello 1     & 4.95 & 16.5  & $1.399 \times 10^{-3}$ & 0.0492 & 0.1096 & 0.0345 & 0.1099 \\
Mello 2     & 4.00 & 8.0   & $2.967 \times 10^{-2}$ & 0.0492 & 0.1096 & 0.0345 & 0.1099 \\
Mello 3     & 4.39 & 4.39  & $9.212 \times 10^{-2}$ & 0.0492 & 0.1096 & 0.0345 & 0.1099 \\
Mello 4     & 18.7 & 6.24  & $2.555 \times 10^{-3}$ & 0.0492 & 0.1096 & 0.0345 & 0.1099 \\
Mello 5     & 14.0 & 0.00  & $8.010 \times 10^{-1}$ & 0.0492 & 0.1096 & 0.0345 & 0.1099 \\
Mello 6     & 29.8 & 0.00  & $3.613 \times 10^{-2}$ & 0.0492 & 0.1096 & 0.0345 & 0.1099 \\
Mello 7     & 73.5 & 0.00  & $2.564 \times 10^{-6}$ & 0.0492 & 0.1096 & 0.0345 & 0.1099 \\
Mello 8     & 0.00 & 9.85  & $3.280 \times 10^{-1}$ & 0.0492 & 0.1096 & 0.0345 & 0.1099 \\
Mello 9     & 0.00 & 15.2  & $1.125 \times 10^{-2}$ & 0.0492 & 0.1096 & 0.0345 & 0.1099 \\
Mello 10    & 0.00 & 32.3  & $1.626 \times 10^{-6}$ & 0.0492 & 0.1096 & 0.0345 & 0.1099 \\
cheR        & 5    & 10    & $20.10$                & 0.02 & 0.5 & 100 & $10^6$ \\
Tar(EEEE)   & 5    & 10    & $4.540 \times 10^{-5}$ & 0.02 & 0.5 & 100 & $10^6$ \\
Tar(QEEE)   & 5    & 10    & $3.060 \times 10^{-7}$ & 0.02 & 0.5 & 100 & $10^6$ \\
Tar(QEQE)   & 5    & 10    & $1.520 \times 10^{-8}$ & 0.02 & 0.5 & 100 & $10^6$ \\
Tar(QEQQ)   & 5    & 10    & $1.250 \times 10^{-9}$ & 0.02 & 0.5 & 100 & $10^6$ \\
CheB        & 7.29 & 10.21 & $1.000$                & 0.02 & 0.5 & 100 & $10^6$ \\
\end{tabular}
\end{ruledtabular}
\end{table}

\subsection{Channel capacity}
\label{sec:capacity_background}

Channel capacity is the maximum mutual information that the receptor channel can transmit between the input ligand concentration and the output (activity). Mutual information $I[C;S]$ quantifies the average amount of information obtained about the input $c$ by observing the output $s\in\{0,1\}$, where $s=0$ denotes inactive and $s=1$ denotes active. For each $c$, the conditional distribution is given by Eq.~\ref{MWC}: $p(s{=}1\mid c)=p(c)$ and $p(s{=}0\mid c)=1-p(c)$. Let $p_{\mathrm{in}}(c)$ be an arbitrary input distribution over concentrations and define the output marginal $p(s)=\int p_{\mathrm{in}}(c)\,p(s \mid c)\,dc$. The mutual information (in bits) is given as

\begin{equation}
I[C;S]=\sum_{s\in\{0,1\}}\int p_{\mathrm{in}}(c)\,p(s\mid c)\log_2\!\left(\frac{p(s\mid c)}{p(s)}\right)\,dc.
\end{equation}
Since the channel has a binary output, $I[C;S]$ is between $0$ and $1$ bits. $I[C;S]=0$ bits indicates that $c$ and $s$ are independent, while $I[C;S]=1$ bit indicates that the output is deterministic. Noise in the sensor or environment decreases mutual information. Channel capacity is the maximum mutual information obtained by choosing the most informative input distribution. Mathematically, it is the supremum over input distributions, given as
\begin{equation}
\bm{C}=\sup_{p_{\mathrm{in}}(c)} I[C;S].
\end{equation}
Biologically, one must argue that the sensor changes its input concentration via movement in order to saturate its information processing capabilities.

\section{Methods}


To evaluate how well the receptor cluster can sense ligand concentration, its sensor properties are quantified by three performance metrics: channel capacity, dynamic range, and effective Hill coefficient. These metrics are computed from the MWC activity curve $p(c)$ while sweeping seven model parameters: the allosteric constant $L_0$, the four state-dependent dissociation constants $K_d^{(I),1}$, $K_d^{(A),1}$, $K_d^{(I),2}$, $K_d^{(A),2}$, and the Tar/Tsr receptor copy numbers $N_1$ and $N_2$.

For each strain operating point, we perform a seven-dimensional parameter sweep by constructing a grid centered around it (Table \ref{tab:1}). Each parameter is varied independently around its reference value while all others are held fixed, with logarithmic spacing used for the allosteric and dissociation constants and linear spacing used for the receptor copy numbers. This procedure defines a discrete seven-dimensional hyper-rectangle surrounding the strain operating point, on which all three performance metrics are evaluated.

To analyze and present the results of the sweep, we extract lower-dimensional representations from the seven-dimensional parameter sweep in the Supplementary Information Sec. S3. Two-parameter slices are visualized as heatmaps, in which two parameters span the axes while the remaining five are fixed at the strain operating point (Table~\ref{tab:1}). The twenty strain operating points are overlaid on these heatmaps as reference markers to indicate where the strains sit in the seven-dimensional parameter sweep. For visualization purposes, heatmap color scales are rescaled using percentile clipping (2nd–98th percentile); all numerical analyses are performed on the unclipped data. One-dimensional slices are additionally used to generate channel capacity, dynamic range, and effective Hill coefficient curves along individual parameter directions.

\subsection{Channel capacity}

Channel capacity $\bm{C}$ is numerically calculated via the Blahut--Arimoto algorithm \cite{blahut1972computation,arimoto1972algorithm}, which iteratively reweights the input distribution to monotonically increase $I[C;S]$ until convergence. In addition to the capacity value $\bm{C}^{*}$, the algorithm also outputs the optimal input distribution $p_{\mathrm{in}}^{*}(c)$ and the optimal output occupancy $p^{*}(s)$. $p_{\mathrm{in}}^*(c)$ is the distribution of ligand concentrations that would realize that capacity, and $p^*(s)$ shows the fraction of time that the receptor ends up in the inactive versus active state at capacity.

We model receptor-cluster sensing as a discrete memoryless channel whose input is ligand concentration and whose output is the binary activity state of the cluster. The concentration axis is represented by a logarithmic grid $\{c_j\}_{j=1}^{M}$ of $M$ bins chosen to sweep over concentrations that showcase both the plateaus and the steep changes in activity curves of Fig.~\ref{MWC curve}. Refining the logarithmic grid $\{c_j\}$ left qualitative conclusions unchanged. For each bin $c_j$, the output $s\in\{0,1\}$ is distributed according to the MWC response (Eq.~\ref{MWC}):
\begin{equation}
\begin{aligned}
P(s=1 \mid c_j) &= p(c_j),\\
P(s=0 \mid c_j) &= 1 - p(c_j).
\end{aligned}
\end{equation}
Let $p_{\mathrm{in}}(c_j)$ denote the prior probability that the input lies in bin $j$. The induced output marginal is $p(s)=\sum_{j=1}^{M} p_{\mathrm{in}}(c_j)\,P(s\mid c_j)$.

The mutual information (in bits) between the input concentration $C$ and the output state $S$ on this grid is
\begin{equation}
I[C;S]=\sum_{j=1}^{M} p_{\mathrm{in}}(c_j)\sum_{s\in\{0,1\}} P(s\mid c_j)\log_2\!\left(\frac{P(s\mid c_j)}{p(s)}\right).
\end{equation}
The static channel capacity is the maximum of $I[C;S]$ over all input priors on the grid,
\begin{equation}
\bm{C}=\max_{p_{\mathrm{in}}(c_j)} I[C;S],
\end{equation}
and because the output is binary one always has $0\le \bm{C}\le 1$ bit per independent observation.

To compute $\bm{C}$ and a maximizing prior, we use the Blahut--Arimoto algorithm for fixed discrete channels. Starting from any initial prior $p_{\mathrm{in}}^{(0)}(c_j)$ with $\sum_{j=1}^{M} p_{\mathrm{in}}^{(0)}(c_j)=1$, define at iteration $t$ the output marginal
\begin{equation}
p^{(t)}(s)=\sum_{j=1}^{M} p_{\mathrm{in}}^{(t)}(c_j)\,P(s\mid c_j),\qquad s\in\{0,1\}.
\end{equation}

For compactness, define the per-input ``row score''
\begin{equation}
\ell_j^{(t)}=\sum_{s\in\{0,1\}} P(s\mid c_j)\,
\ln\!\left(\frac{P(s\mid c_j)}{p^{(t)}(s)}\right).
\end{equation}
The prior is then updated by normalized reweighting:
\begin{equation}
p_{\text{in}}^{(t+1)}(c_j)
=
\frac{
p_{\text{in}}^{(t)}(c_j)\,\exp\!\big(\ell_j^{(t)}\big)
}{
\sum_{k} p_{\text{in}}^{(t)}(c_k)\,\exp\!\big(\ell_k^{(t)}\big)
}\,.
\end{equation}
which guarantees $\sum_{j}p_{\mathrm{in}}^{(t+1)}(c_j)=1$. These iterations monotonically increase $I[C;S]$ and converge to a capacity-achieving prior $p_{\mathrm{in}}^*(c_j)$. We report the capacity $\bm{C}^{*}=I_{p_{\mathrm{in}}^{*}}[C;S]$ (converted to bits via $\log_{2}$), together with the optimal input distribution $p_{\mathrm{in}}^*(c_j)$ and the induced optimal output occupancy
\begin{equation}
p^*(s)=\sum_{j=1}^{M} p_{\mathrm{in}}^*(c_j)\,P(s\mid c_j),\qquad s\in\{0,1\}.
\end{equation}
Because the output is binary, the maximizing prior typically concentrates probability near concentrations whose outputs are the most distinct.

\subsection{Dynamic range}

Dynamic range is defined as the change in mean activity between the basal (no input) and saturated (very high input) states. We have
\begin{eqnarray}
p_0 &=& p(0)=\frac{1}{1+L_0}
\label{p0}
\end{eqnarray}
and
\begin{eqnarray}
p_\infty &=& \lim_{c\to\infty} p(c) \nonumber \\
&=& \frac{1}{1+L_0
\left(\frac{K_{d}^{(A),1}}{K_{d}^{(I),1}}\right)^{N_1}
\left(\frac{K_{d}^{(A),2}}{K_{d}^{(I),2}}\right)^{N_2}}
\label{pinf}
\end{eqnarray}
with
\begin{equation}
\mathrm{DR} = \bigl|\,p_\infty - p_0\,\bigr|.
\label{dynamic range}
\end{equation}

\subsection{Effective Hill coefficient}

Cooperative ligand-receptor binding interactions can be fit to the Hill equation:
\begin{equation}
 p(c) = \frac{(c/K_d)^{n_{\mathrm{eff}}}}{1+(c/K_d)^{n_{\mathrm{eff}}}}. 
\end{equation}
Physically, the effective Hill coefficient $n_{\mathrm{eff}}$ quantifies how abruptly the output changes as $c$ crosses its midpoint, indicating that the binding sites act in concert (cooperativity). For attractants, the activity response $p(c)$ is decreasing; we report $\lvert n_{\mathrm{eff}}\rvert$ as the steepness magnitude. Negative cooperativity comes from $|n_{\mathrm{eff}}|<1$, while positive cooperativity comes from $|n_{\mathrm{eff}}|>1$. Higher deviations of $|n_{\mathrm{eff}}|$ from $1$ mean a more switch-like response (small input changes cause large output changes near the concentration that leads to middling $p(c)$), but may narrow the dynamic range of the sensor.

To find the effective Hill coefficient, define the transition output level as the midpoint
\begin{equation}
\label{eq:midpoint}
p_{\mathrm{mid}} \equiv \frac{p_0+p_\infty}{2},
\end{equation}
and define $c^*$ as the unique concentration that satisfies
\begin{equation}
\label{eq:cstar}
p(c^*) = p_{\mathrm{mid}},
\end{equation}
and define the normalized activity
\begin{equation}
\label{eq:ac-def}
a(c)\equiv \frac{p(c)-p_0}{p_\infty-p_0}.
\end{equation}
The effective Hill coefficient is \cite{marzen2013statistical}
\begin{equation}
\label{eq:neff-def}
n_{\mathrm{eff}}
=2\,\left.\frac{d}{d\ln c}\ln a(c)\right|_{c=c^*}.
\end{equation}
As shown in the Supplementary Information Sec.~S2,
\begin{eqnarray}
\label{eq:neff-compact}
n_{\mathrm{eff}}
&=&
-\,4\,
\frac{c^*\,p_{\mathrm{mid}}(1-p_{\mathrm{mid}})}{\,p_\infty-p_0\,}
\nonumber\\
&& \times \Bigg[
N_1\!\left(\frac{1}{K_d^{(I),1}+c^*}-\frac{1}{K_d^{(A),1}+c^*}\right)
\nonumber\\
&& \quad + N_2\!\left(\frac{1}{K_d^{(I),2}+c^*}-\frac{1}{K_d^{(A),2}+c^*}\right)
\Bigg].
\end{eqnarray}

\subsection{Numerical gradient calculation}

To interpret the seven-dimensional sweep in a way that is comparable across strains, we summarize the results using two complementary quantities. The value of each metric at each strain's operating point is compared to the global maximum attained over the seven-dimensional region swept, and the $L_2$ norm of the local gradient at the strain parameters is compared to the maximum gradient $L_2$ norm observed. Together, these comparisons assess whether the strain operating points are positioned near globally optimal sensing performance and whether they lie in locally flat or steep regions of the performance landscape.

For each performance metric $M \in \{\bm{C},\mathrm{DR},n_{\mathrm{eff}}\}$, we quantify local flatness at the strain operating point by the Euclidean ($L_2$) norm of the gradient in a mixed coordinate system,
\begin{eqnarray}
\mathbf{u} &=& \bigl(\log_{10}L_0,\log_{10}K_d^{(I),1},\log_{10}K_d^{(A),1},\log_{10}K_d^{(I),2} \nonumber \\
&& ,\log_{10}K_d^{(A),2},N_1,N_2\bigr),
\end{eqnarray}
where the five biochemical parameters are treated on a logarithmic scale and receptor copy numbers are treated linearly. 
In the vector of strain parameters $\mathbf{u}$, partial derivatives are estimated numerically using fixed-step finite differences. 
For a logarithmic parameter $x \in \{L_0,K_d^{(s),i}\}$, we approximate the derivative with respect to $\log_{10}x$ by
\begin{equation}
\frac{\partial M}{\partial \log_{10}x}
\approx
\frac{M(x\,10^{\delta})-M(x\,10^{-\delta})}{2\delta},
\qquad \delta = 0.01,
\end{equation}
and for receptor copy numbers $N\in\{N_1,N_2\}$ we use
\begin{equation}
\frac{\partial M}{\partial N}
\approx
\frac{M(N+\Delta)-M(N-\Delta)}{2\Delta},
\qquad \Delta = 1.
\end{equation}
When a symmetric step leaves the swept bounds, we use the corresponding one-sided difference. 
The local gradient norm is then
\begin{equation}
\|\nabla_{\mathbf{u}} M\|_2
=
\left(\sum_{i=1}^{7}\left(\frac{\partial M}{\partial u_i}\right)^2\right)^{1/2}.
\end{equation}

To contextualize the local gradient magnitude, we estimate the maximum gradient norm attainable within the swept seven-dimensional hyper-rectangle. Because evaluating $\|\nabla_{\mathbf{u}} M\|_2$ throughout the entire seven-dimensional grid is computationally expensive, we instead approximate this maximum using a stochastic optimization procedure. Because this procedure samples only a finite number of randomly sampled points, the resulting value is a conservative stochastic estimate of the maximum gradient norm over the swept region. In each round, the procedure draws candidate points uniformly within the current bounding box in $\mathbf{u}$-space, evaluates $\|\nabla_{\mathbf{u}} M\|_2$ at each point, retains the best point, and shrinks the bounding box around it by a fixed factor. We repeat this procedure for five independent random seeds and take the largest value across seeds as our estimate of the global maximum gradient norm over the swept region. Since all extrema reported over the seven-dimensional sweep are taken over a finite, discretized parameter grid, they are conservative estimates of the corresponding continuous extrema. Finally, we report the dimensionless ratio
\begin{equation}
\frac{\|\nabla_{\mathbf{u}} M\|_2\big|_{\mathrm{strain}}}{\max_{\mathrm{swept\ region}} \|\nabla_{\mathbf{u}} M\|_2},
\end{equation}
which quantifies how flat (small ratio) or steep (large ratio) the metric landscape is in the vicinity of each strain operating point. 

\section{Results}

First, in Section \ref{sec:static}, we describe why the channel capacity in the static case is relevant for biological function when bacteria use trajectories of sensor states to understand trajectories of concentrations, in particular, temporal derivatives of concentrations. Next, in Section \ref{sec:relationships}, we describe how static metrics (including the channel capacity thus defined) are related to one another.

\subsection{Why the static case matters}
\label{sec:static}

Let $C_{t':t}$ denote the chain of random variables $C_{t'},C_{t'+1},...,C_{t}$ of concentrations over time, and let $S_{t':t}$ similarly denote the chain of random variables $S_{t'},S_{t'+1},...,S_t$ of sensor states over time. It is often desirable to find the mutual information from the trajectory $I[C_{0:N};S_{0:N}]$, as this is very useful for understanding the ability of sensors to detect temporal gradients and not just static concentrations \cite{tostevin2009mutual,auconi2022gradient}.
We now relate the trajectory mutual information $I[C_{0:N};S_{0:N}]$ to the single-timestep mutual information $I[C;S]$ by repeatedly applying the chain rule for mutual information and adding four additional assumptions: first, that feedback effects can be neglected; second, that the concentration evolves autonomously; third, that the sensor's future state depends only on its past states and the current concentration; and fourth, that the sensor's internal noise does not feed back into future concentrations. We start with the following.
\begin{eqnarray}
I[C_{0:N};S_{0:N}] &=& \sum_t I[C_{0:N};S_t|S_{0:t-1}] \\
&=& \sum_t I[C_t;S_t|S_{0:t-1}]\nonumber\\
&& +I[C_{0:t-1},C_{t+1:N};S_t|C_t,S_{0:t-1}] \\
&=& \sum_t I[C_t;S_t|S_{0:t-1}] \nonumber \\
&& + I[C_{t+1:N};S_t|C_{0:t},S_{0:t-1}]\nonumber \\
&& +I[C_{0:t-1};S_t|C_t,S_{0:t-1}]
\end{eqnarray}
where $I[C_{t+1:N};S_t|C_{0:t},S_{0:t-1}]=0$ if the concentrations evolve autonomously, that is, the concentration evolves independently of the sensor state. For $I[C_{t+1:N};S_t|C_{0:t},S_{0:t-1}]$ to vanish it is also required that, conditional on the concentration history $C_{0:t}$, the sensor's internal noise up to time $t$ is independent of the future concentrations $C_{t+1:N}$; equivalently, future concentrations do not depend on the sensor's internal noise beyond what $C_{0:t}$ already contains. So we have
\begin{eqnarray}
I[C_{0:N};S_{0:N}] &=& \sum_t I[C_t;S_t|S_{0:t-1}] \nonumber \\
&& + I[C_{0:t-1};S_t|C_t,S_{0:t-1}].
\end{eqnarray}
But $s_t$ depends only on $s_{0:t-1}$ and $c_t$, using a typical time-step convention in which the input advances first and the sensor state advances second, and so the second term $I[C_{0:t-1};S_t|C_t,S_{0:t-1}]=0$. This leaves us with the following.
\begin{eqnarray}
   I[C_{0:N};S_{0:N}] &=& \sum_t  I[C_t;S_t|S_{0:t-1}] \\
   &\leq & \sum_t I[C;S] \\
   &=& N I[C;S]
\end{eqnarray}
where $I[C;S]$ is the mutual information based on the joint distribution of concentration and sensor, where stationarity is assumed.

The inequality $I[C_t;S_t|S_{0:t-1}]\leq I[C;S]$ is assumed to be biologically plausible. Note that the inequality need not hold in general. One may naively think that conditioning reduces mutual information, but notable examples in which conditioning increases mutual information occur when the past sensor states are functions of the present concentration and the present sensor state. Then, without the conditioning, the two are independent, and with the conditioning, the two acquire dependence. We are clearly far from this regime with a sensor that, to function, must require the present sensor state to be more reflective of the present concentration than past sensor states. The inequality thus seems reasonable because $C_t$ and $S_t$ lose dependence when conditioned by the past of $S$ as the past sensor states are correlated with both the present concentration and the present sensor state. The biological sensors being considered require that the present sensor state understands something about the present concentration to enable function. 

We have dealt with discrete-time measurements. In continuous-time formulations, measurements are made every $\tau$, where $\tau$ is a perceptual timescale below which the bacterium cannot read out activity. This is plausibly the relaxation timescale of sensor readings. We take $\tau$ to be the pathway response time measured in single cells, $\tau = 0.22 \pm 0.01\ \mathrm{s}$ \cite{mattingly2021escherichia}. This timescale is set by the dephosphorylation of the response regulator CheY-P, which time-averages the kinase activity \cite{sartori2011noise}.
The cell does not see activity fluctuations faster than $\tau$ because the signaling pathway downstream of the receptors has a finite response time $\tau$. The cell only gets a new reading about every $\tau$. Thus, $N=T/\tau$, and
\begin{equation}
I[C_{0:N};S_{0:N}] \leq \frac{T}{\tau} I[C;S].
\end{equation}
In other words,
\begin{equation}
\lim_{T\rightarrow \infty} \frac{1}{T}I[C_{0:N};S_{0:N}] \leq \frac{1}{\tau}I[C;S]
\end{equation}
implying that the static mutual information, divided by the perceptual timescale $\tau$, upper-bounds the trajectory mutual information rate. We can take this further by noting that the channel capacity of the static channel upper-bounds the mutual information, even if channel capacity is not reached, meaning that
\begin{equation}
\lim_{T\rightarrow \infty} \frac{1}{T}I[C_{0:N};S_{0:N}] \leq \frac{\bm{C}}{\tau}.
\end{equation}
Thus, success in a trajectory sense implies success in a static channel capacity sense, and if there is no success in a static channel capacity sense, there cannot be success in a trajectory sense. The faster the perceptual timescale, the weaker the bound. In the idealized limit $\tau\to 0$ the right-hand side diverges, so the bound carries no content for a sensor that can be read out with unbounded bandwidth, while it is informative precisely because the biological readout time is finite. With $\tau=0.22\ \mathrm{s}$, the wild-type capacity ceilings $\bm{C}_{\max}(p_0)\approx 0.20$--$0.32\ \mathrm{bits}$ derived in Sec.~\ref{sec:capacity_results} (Table~\ref{tab:cmax_ceiling}) correspond to rate ceilings of roughly $0.9$--$1.5\ \mathrm{bits/s}$, and strains with baseline activity near one have ceilings up to roughly $4.5\ \mathrm{bits/s}$. The bound scales as $1/\tau$: for $\tau \approx 1/60\ \mathrm{s}$ (intrinsic kinase response time) the wild-type ceilings are $12$--$19\ \mathrm{bits/s}$, whereas for $\tau \approx 1\ \mathrm{s}$ (roughly the duration of a run) they are $0.20$--$0.32\ \mathrm{bits/s}$. In either case these ceilings remain one to three orders of magnitude above information rates measured in shallow gradients ($\sim 10^{-2}\ \mathrm{bits/s}$ \cite{mattingly2021escherichia}), so the conclusion that the static channel capacity is not the limiting factor does not depend on the precise choice of $\tau$.

This proof also relies on the autonomy of the concentration time series and the conditionally Markovian nature of the sensor state. If the sensor state is defined expansively enough, the latter is true, so that the past sensor states shield the future sensor states from all but the current concentration. Note that this is not a statement about whether or not the sensor is understanding temporal derivatives; it well may, but it may do so through a more biologically plausible scheme by which its future state only depends on its past state and the current concentration, over time accumulating information about trajectories as do reservoir computers. (To see this, one can unfurl a recurrence relation like $s_{t+1}=f(s_t,c_{t+1})$ into the explicit relationship $s_{t+1}=f(f(f(...f(s_0,c_1),c_2),...,c_{t+1}))$, which means that a single sensory state can have information about the trajectory of concentrations through this scheme.) The former is approximately true in that even though there is a feedback loop in which the actions of a single bacterium affect what it sees, when there are a large number of other bacteria, this effect may approximately wash out.

Even though the static channel capacity, divided by the perceptual timescale $\tau$, upper-bounds the trajectory mutual information rate, dynamic considerations can be folded into this static channel capacity. The inequality merely says that $\rho(c_t)p(s_t|c_t)$ is a relevant distribution for upper-bounding the trajectory mutual information rate. It does not say that adaptation and temporal responses cannot alter that joint probability distribution in order to alter the so-called static channel capacity.
We simply work in the regime described in Sec.~\ref{sec:mwc}, in which concentrations are low and encounters with attractant are brief compared to the adaptation time, but future work should relax this assumption.

\subsection{Relationships amongst static metrics}
\label{sec:relationships}

To quantify the relationships among the three static metrics in the seven-dimensional parameter sweep, we computed pairwise correlation coefficients $R$ over all sampled grid points (Table~\ref{tab:table_corr_R}). All three metrics are positively correlated, as was true for populations of MWC receptors \cite{martins2011trade}. The channel capacity is strongly correlated with the dynamic range ($R=0.986$) and moderately with the effective Hill coefficient ($R=0.556$); the effective Hill coefficient and the dynamic range are also moderately correlated ($R=0.470$). The mechanistic origin of the strong correlation $\bm{C}$--$\mathrm{DR}$ is discussed in the channel capacity subsection below: both metrics reduce to essentially one-parameter functions of the baseline activity $p_0$ in this regime.

\begin{table}[ht]
\centering
\begin{tabular}{lccc}
\toprule
        & \textbf{$\bm{C}$} & \textbf{$|n_{\mathrm{eff}}|$} & \textbf{$\mathrm{DR}$} \\
$\bm{C}$                          & 1     & 0.556 & 0.986 \\
$|n_{\mathrm{eff}}|$         & 0.556 & 1     & 0.470 \\
$\mathrm{DR}$                & 0.986 & 0.470 & 1     \\
\end{tabular}
\caption{Pairwise Pearson correlation coefficient between the three static performance metrics, computed over all $9.56 \times 10^{6}$ valid grid points in the 7D parameter sweep.}
\label{tab:table_corr_R}
\end{table}

For clarity, this section displays a representative subset of heatmaps with strain operating points overlaid. See Supplementary Information Sec.~S3 for the full set of heatmaps across all twenty strains.

\subsubsection{Channel capacity}
\label{sec:capacity_results}

The receptor output is binary, so a classical theorem (Supplementary Information Sec.~S1) guarantees that the capacity-achieving input distribution can always be taken to place all its weight on at most two ligand concentrations, confirmed by numerical simulations. The output activity $s$ is Bernoulli with parameter $p(c)$, and as $c$ varies this parameter ranges over the interval $[p_\infty, p_0]$, where $p_0 = 1/(1+L_0)$ is the baseline activity at the strain operating point and $p_\infty$ is given by Eq.~\eqref{pinf}. The resulting two-input binary channel has $P(s{=}1 \mid c_{\text{low}}) = p_0$ and $P(s{=}1 \mid c_{\text{high}}) = p_\infty$. In the limit $p_\infty \to 0$ the high-concentration input becomes noiseless and the channel is a Z-channel with a low-concentration crossover probability $1 - p_0$, whose capacity is closed form.
\begin{equation}
\label{eq:Cmax_Astar}
\bm{C}_{\max}(p_0) = \log_2\!\left(\,1 + p_0\,(1-p_0)^{(1-p_0)/p_0}\,\right).
\end{equation}
This is an upper limit on the receptor channel capacity at any operating point with baseline activity $p_0$.

The Z-channel limit $p_\infty \to 0$ is a good approximation at every strain operating point considered here, so each strain sits close to the Z-channel ceiling in Eq.~\eqref{eq:Cmax_Astar} (Table~\ref{tab:cmax_ceiling}). For example, WT-EW has $L_0 = 2$, $N_1 = 6$, and $N_2 = 12$; substituting these into Eq.~\eqref{pinf} gives a denominator of the order of $10^{56}$, and therefore $p_\infty \lesssim 10^{-56}$. This also explains the strong correlation between $\bm{C}$ and $\mathrm{DR}$ in Table~\ref{tab:table_corr_R}: because $p_\infty \ll p_0$ in every strain, the dynamic range $\mathrm{DR} = |p_\infty - p_0| \approx p_0$, so both $\bm{C}$ and $\mathrm{DR}$ depend primarily on $p_0$.

\begin{table}[h!]
\centering
\small
\setlength{\tabcolsep}{4pt}
\begin{ruledtabular}
\begin{tabular}{lcccc}
Strain & $p_0$ & $\bm{C}$ (bits) & $\bm{C}_{\max}(p_0)$ & $\bm{C}/\bm{C}_{\max}(p_0)$ \\
\hline
WT-K        & 0.500 & 0.321 & 0.322 & 0.997 \\
WT-EW       & 0.333 & 0.199 & 0.199 & 0.996 \\
WT-H        & 0.333 & 0.199 & 0.199 & 0.996 \\
WT-C        & 0.345 & 0.207 & 0.207 & 0.997 \\
\hline
Mello 1     & 0.999 & 0.989 & 0.992 & 0.997 \\
Mello 2     & 0.971 & 0.902 & 0.906 & 0.995 \\
Mello 3     & 0.916 & 0.776 & 0.790 & 0.982 \\
Mello 4     & 0.998 & 0.984 & 0.987 & 0.997 \\
Mello 5     & 0.555 & 0.366 & 0.368 & 0.995 \\
Mello 6     & 0.965 & 0.890 & 0.891 & 0.998 \\
Mello 7     & 1.000 & 0.999 & 1.000 & 0.999 \\
Mello 8     & 0.753 & 0.559 & 0.562 & 0.996 \\
Mello 9     & 0.989 & 0.953 & 0.956 & 0.997 \\
Mello 10    & 1.000 & 0.999 & 1.000 & 0.999 \\
cheR        & 0.047 & 0.025 & 0.025 & 0.987 \\
Tar(EEEE)   & 1.000 & 0.995 & 1.000 & 0.995 \\
Tar(QEEE)   & 1.000 & 0.998 & 1.000 & 0.998 \\
Tar(QEQE)   & 1.000 & 0.998 & 1.000 & 0.998 \\
Tar(QEQQ)   & 1.000 & 0.998 & 1.000 & 0.998 \\
CheB        & 0.500 & 0.321 & 0.322 & 0.997 \\
\end{tabular}
\end{ruledtabular}
\caption{Channel capacity $\bm{C}$ compared with the Z-channel ceiling $\bm{C}_{\max}(p_0)$ from Eq.~\eqref{eq:Cmax_Astar} at each strain operating point. The first four rows are the wild-type strains; the remaining sixteen are lab-grown mutants. Every operating point achieves $\bm{C}/\bm{C}_{\max}(p_0) \geq 0.98$, demonstrating that the receptor channel saturates the Z-channel ceiling set by its own baseline activity $p_0$. The residual sub-percent gap arises because the finite log-spaced concentration grid used by the Blahut--Arimoto algorithm does not reach $p_\infty$ exactly. The comparison with maximum possible achievable channel capacity in the 7D sweep is simple; that maximal channel capacity is simply $1$ bit.}
\label{tab:cmax_ceiling}
\end{table}

\begin{figure}[h!]
    \centering
    \includegraphics[width=\linewidth]{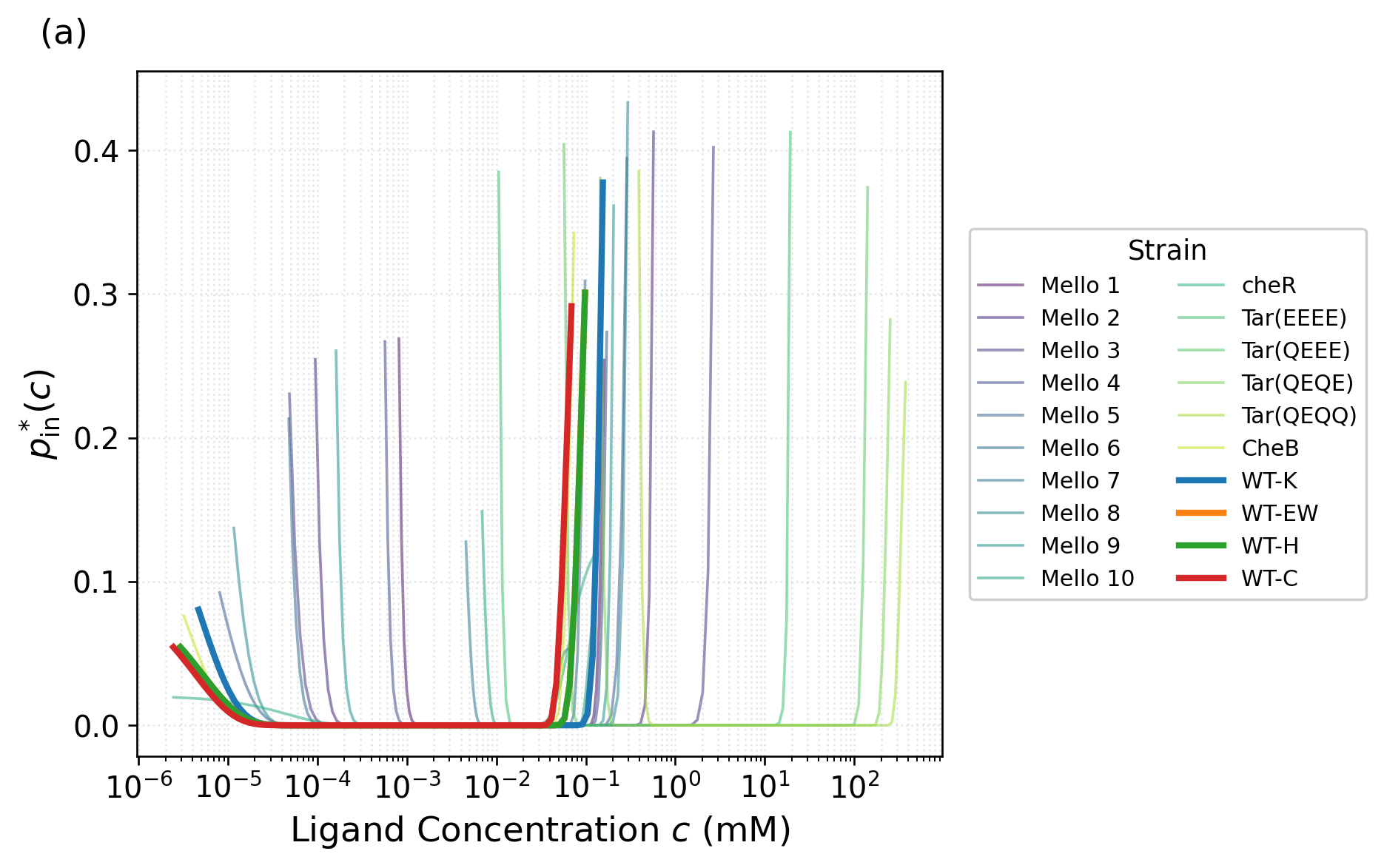}
    \includegraphics[width=\linewidth]{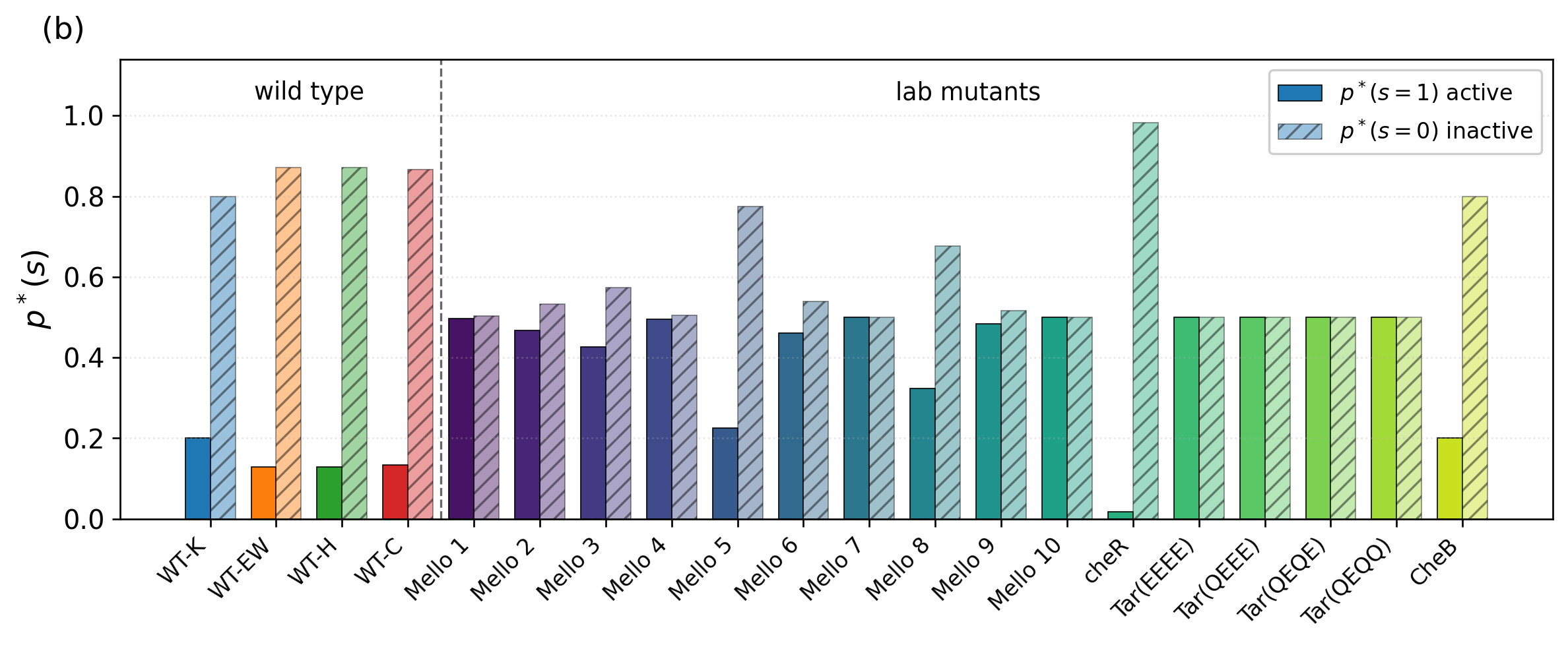}
    \caption{(a) Capacity-achieving input distribution $p_{\mathrm{in}}^*(c)$ versus ligand concentration $c$ at all twenty strain operating points in Table~\ref{data}, obtained via the Blahut--Arimoto algorithm. (b) Induced optimal output occupancy $p^*(s)$ between the active and inactive states at each strain operating point; the four wild-type strains are separated from the sixteen lab-grown mutants by the dashed line. In both panels, wild-type strains are shown in saturated colors and lab-grown mutants in lighter shades.}
    \label{global}
\end{figure}

\begin{figure}[h!]
    \centering
    \includegraphics[width=\linewidth]{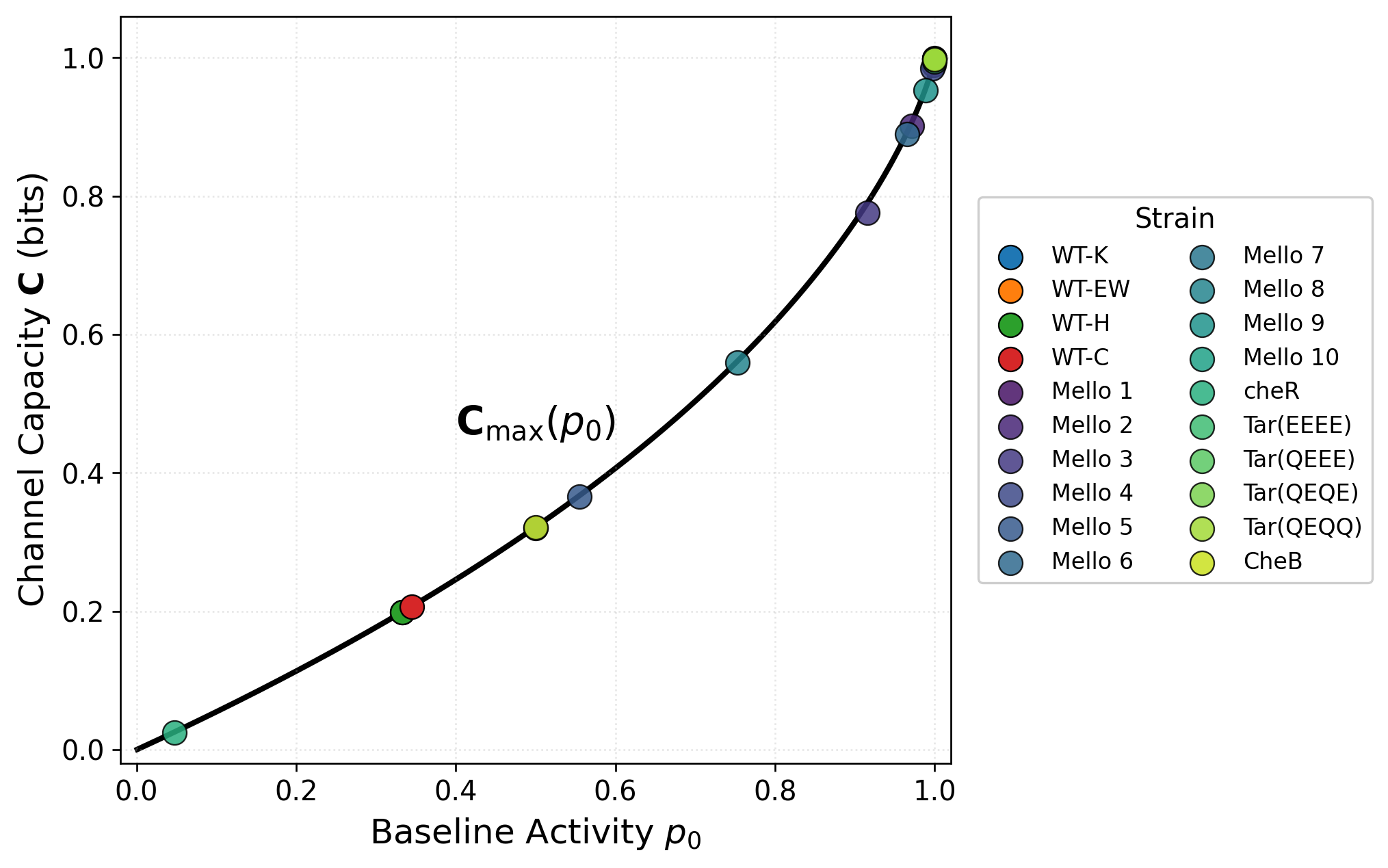}
    \caption{Channel capacity $\bm{C}$ at each of the twenty strain operating points in Table~\ref{data}, plotted against the baseline activity $p_0 = 1/(1+L_0)$. The solid curve is the Z-channel ceiling $\bm{C}_{\max}(p_0)$ of Eq.~\eqref{eq:Cmax_Astar}, which upper-bounds the MWC receptor channel capacity in the limit $p_\infty \to 0$. Every strain sits very close to the ceiling (Table~\ref{tab:cmax_ceiling}). Wild-type strains are shown in saturated colors; lab-grown mutants in lighter shades.}
    \label{cmax_curve}
\end{figure}


\begin{table}[h!]
\centering
\small
\setlength{\tabcolsep}{4pt}
\begin{tabular}{lccc}
\toprule
\textbf{Strain} & $\dfrac{\|\nabla \bm{C}\|_2}{\|\nabla \bm{C}\|_\text{max}}$ & $\dfrac{\|\nabla \mathrm{DR}\|_2}{\|\nabla \mathrm{DR}\|_\text{max}}$ & $\dfrac{\|\nabla |n_{\mathrm{eff}}|\|_2}{\|\nabla |n_{\mathrm{eff}}|\|_\text{max}}$ \\[6pt]
WT-K        & 0.039 & 0.040 & 0.002 \\
WT-EW       & 0.040 & 0.040 & 0.002 \\
WT-H        & 0.040 & 0.040 & 0.002 \\
WT-C        & 0.040 & 0.040 & 0.002 \\
\hline
Mello 1     & 0.003 & 0.001 & 0.021 \\
Mello 2     & 0.026 & 0.007 & 0.015 \\
Mello 3     & 0.062 & 0.026 & 0.010 \\
Mello 4     & 0.005 & 0.001 & 0.025 \\
Mello 5     & 0.090 & 0.041 & 0.011 \\
Mello 6     & 0.018 & 0.006 & 0.020 \\
Mello 7     & 0.001 & 0.000 & 0.084 \\
Mello 8     & 0.054 & 0.032 & 0.007 \\
Mello 9     & 0.017 & 0.006 & 0.015 \\
Mello 10    & 0.002 & 0.000 & 0.069 \\
cheR        & 0.039 & 0.040 & 0.002 \\
Tar(EEEE)   & 0.001 & 0.000 & 0.013 \\
Tar(QEEE)   & 0.000 & 0.000 & 0.007 \\
Tar(QEQE)   & 0.001 & 0.000 & 0.023 \\
Tar(QEQQ)   & 0.001 & 0.000 & 0.023 \\
CheB        & 0.040 & 0.040 & 0.002 \\
\end{tabular}
\caption{Local flatness of channel capacity, dynamic range, and effective Hill coefficient at each strain operating point, expressed as the $L_2$ norm of the numerical gradient of each performance metric divided by its maximum over the 7D parameter sweep. Global maxima are $\|\nabla \bm{C}\|_\text{max} = 6.64$, $\|\nabla \mathrm{DR}\|_\text{max} = 6.66$, and $\|\nabla |n_{\mathrm{eff}}|\|_\text{max} = 215.23$. The first four rows are the wild-type strains; the remaining sixteen are lab-grown mutants. All gradients are estimated numerically on the sweep grid using centered finite differences with logarithmic spacing for $L_0$ and $K_d$ and linear spacing for $N_1, N_2$.}
\label{tab:gradnorm_allmetrics}
\end{table}


Therefore, the qualitative parameter dependencies of $\bm{C}$ can be read off directly from the dynamic-range expression, and we defer a detailed discussion of these trends to the following subsection.

\subsubsection{Dynamic range}

Dynamic range quantifies how far the receptor output can move between the basal and saturated limits. From Eq. \eqref{dynamic range} and as visually corroborated by the heatmaps, dynamic range is governed by at least three qualitative trends. First, \(\mathrm{DR}\) is suppressed when
\(K_{d}^{(A),i}\approx K_{d}^{(I),i}\), because \(K_{d}^{(A),i}/ K_{d}^{(I),i}\approx1\) and \(p_\infty\approx p_0\). Second, in the attractant regime \(K_{d}^{(I),i}<K_{d}^{(A),i}\), increasing the
separation between active- and inactive-state affinities and/or increasing receptor copy numbers \((N_1,N_2)\) amplifies the product term and pushes \(p_\infty\) farther from \(p_0\), thereby increasing
\(\mathrm{DR}\) toward its upper bound. Third, \(\mathrm{DR}\) can remain small even when \(K_{d}^{(A),i}\neq K_{d}^{(I),i}\) due to extreme allosteric bias. For \(L_0\ll 1\), \(p_0\approx 1\) and typically \(p_\infty\approx 1\) unless the product term is large enough to overcome the small prefactor \(L_0\); for \(L_0\gg 1\), \(p_0\approx 0\) and typically \(p_\infty\approx 0\) as well. 

Table~\ref{tab:tableA_DR} shows that most lab-grown mutants achieve dynamic range values close to the global maximum of $1$, while the four wild-type strains sit at $\mathrm{DR} \approx 0.33$--$0.50$. The normalized gradient norms in Table~\ref{tab:gradnorm_allmetrics} confirm that every strain operating point is in a locally flat region of the dynamic-range landscape ($\|\nabla \mathrm{DR}\|_2 / \|\nabla \mathrm{DR}\|_\text{max} \leq 0.041$ at all twenty strains).

\begin{table}[h!]
\centering
\begin{tabular}{lccc}
\toprule
\textbf{Strain} & \textbf{$\mathrm{DR}_\text{strain}$} & \textbf{$\mathrm{DR}_\text{max}$} & \textbf{$\mathrm{DR}_\text{strain}$ / $\mathrm{DR}_\text{max}$} \\
WT-K        & 0.50 & 1.00 & 0.50 \\
WT-EW       & 0.33 & 1.00 & 0.33 \\
WT-H        & 0.33 & 1.00 & 0.33 \\
WT-C        & 0.34 & 1.00 & 0.34 \\
\hline
Mello 1     & 1.00 & 1.00 & 1.00 \\
Mello 2     & 0.97 & 1.00 & 0.97 \\
Mello 3     & 0.91 & 1.00 & 0.91 \\
Mello 4     & 1.00 & 1.00 & 1.00 \\
Mello 5     & 0.56 & 1.00 & 0.56 \\
Mello 6     & 0.97 & 1.00 & 0.97 \\
Mello 7     & 1.00 & 1.00 & 1.00 \\
Mello 8     & 0.75 & 1.00 & 0.75 \\
Mello 9     & 0.99 & 1.00 & 0.99 \\
Mello 10    & 1.00 & 1.00 & 1.00 \\
cheR        & 0.05 & 1.00 & 0.05 \\
Tar(EEEE)   & 1.00 & 1.00 & 1.00 \\
Tar(QEEE)   & 1.00 & 1.00 & 1.00 \\
Tar(QEQE)   & 1.00 & 1.00 & 1.00 \\
Tar(QEQQ)   & 1.00 & 1.00 & 1.00 \\
CheB        & 0.50 & 1.00 & 0.50 \\
\end{tabular}
\caption{Dynamic range $\mathrm{DR}_{\text{strain}}$ for each strain operating point, compared with the maximum $\mathrm{DR}_{\max}$ over the 7D parameter sweep. The first four rows are the wild-type strains; the remaining sixteen are lab-grown mutants.}

\label{tab:tableA_DR}
\end{table}



\subsubsection{Effective Hill coefficient}

Effective Hill coefficient quantifies how sharply the receptor output switches between its basal and saturated limits around the midpoint concentration. From Eq.~\eqref{eq:neff-compact} (and consistent with the heatmaps), the effective Hill coefficient is governed by several qualitative trends. First, $|n_{\mathrm{eff}}|$ increases with receptor copy number: because $N_1$ and $N_2$ enter linearly in the sensitivity term, larger Tar/Tsr cluster sizes yield a more switch-like transition near the midpoint. 
Second, $|n_{\mathrm{eff}}|$ increases as the difference between inactive- and active-state dissociation constants increases. When $K_{d}^{(I),i}\approx K_{d}^{(A),i}$ the difference terms $\left(\frac{1}{K_{d}^{(I),i}+c^*}-\frac{1}{K_{d}^{(A),i}+c^*}\right)$ vanish, producing a shallow response and thus small $|n_{\mathrm{eff}}|$; conversely, larger $K_{d}^{(A),i}/K_{d}^{(I),i}$ amplifies these terms and steepens the curve. Third, the prefactor $-4\,\frac{c^*\,p_{\mathrm{mid}}(1-p_{\mathrm{mid}})}{p_\infty-p_0}$ shows that endpoint saturation matters: if the activity curve does not meaningfully span distinct low- and high-concentration plateaus ($p_\infty\approx p_0$), then $|n_{\mathrm{eff}}|$ becomes ill-conditioned, and can be undefined when the response is effectively flat.

The strain effective Hill coefficients in Table~\ref{tab:tableA_nH} are all far from the global maximum $|n_{\mathrm{eff}}|_\text{max}=62.82$, with ratios ranging from $0.02$ at wild-type operating points to $0.16$ at Mello 7. Table~\ref{tab:gradnorm_allmetrics} shows that the normalized gradient of $|n_{\mathrm{eff}}|$ is small at every strain ($\leq 0.084$), so each strain sits in a locally flat region of the sweep, but the strain values themselves are far below the global maximum. From this we conclude that the effective Hill coefficient is not globally maximized at any strain operating point in Table~\ref{data}.


\begin{table}[h!]
\centering
\begin{tabular}{lccc}
\toprule
\textbf{Strain} & \textbf{$\lvert n_{\mathrm{eff}}\rvert_\text{strain}$} & \textbf{$\lvert n_{\mathrm{eff}}\rvert_\text{max}$} & \textbf{$\lvert n_{\mathrm{eff}}\rvert_\text{strain}$ / $\lvert n_{\mathrm{eff}}\rvert_\text{max}$} \\
WT-K        & 1.46 & 62.82 & 0.02 \\
WT-EW       & 1.41 & 62.82 & 0.02 \\
WT-H        & 1.41 & 62.82 & 0.02 \\
WT-C        & 1.43 & 62.82 & 0.02 \\
\hline
Mello 1     & 4.61 & 62.82 & 0.07 \\
Mello 2     & 2.57 & 62.82 & 0.04 \\
Mello 3     & 1.91 & 62.82 & 0.03 \\
Mello 4     & 4.31 & 62.82 & 0.07 \\
Mello 5     & 1.52 & 62.82 & 0.02 \\
Mello 6     & 2.99 & 62.82 & 0.05 \\
Mello 7     & 9.97 & 62.82 & 0.16 \\
Mello 8     & 1.71 & 62.82 & 0.03 \\
Mello 9     & 3.32 & 62.82 & 0.05 \\
Mello 10    & 8.37 & 62.82 & 0.13 \\
cheR        & 1.30 & 62.82 & 0.02 \\
Tar(EEEE)   & 3.19 & 62.82 & 0.05 \\
Tar(QEEE)   & 1.26 & 62.82 & 0.02 \\
Tar(QEQE)   & 1.93 & 62.82 & 0.03 \\
Tar(QEQQ)   & 3.63 & 62.82 & 0.06 \\
CheB        & 1.52 & 62.82 & 0.02 \\
\end{tabular}
\caption{Effective Hill coefficient $\lvert n_{\mathrm{eff}}\rvert_\text{strain}$ for each strain operating point, compared with the maximum $\lvert n_{\mathrm{eff}}\rvert_\text{max}$ over the 7D parameter sweep. The first four rows are the wild-type strains; the remaining sixteen are lab-grown mutants.}
\label{tab:tableA_nH}
\end{table}




\section{Conclusions}

This study examined the static sensing limits of mixed Tar/Tsr chemoreceptor clusters in \textit{E.\ coli} using a heterogeneous MWC framework. By performing a seven-dimensional parameter sweep and computing three performance metrics---channel capacity, effective Hill coefficient, and dynamic range---we mapped how receptor composition and state-dependent affinities shape the information available at the receptor output level.


WT strains sit in locally flat regions of all three landscapes; their operating points lie below the global maxima for channel capacity, dynamic range, and effective Hill coefficient. Engineered mutants achieve high channel capacity and dynamic range because their biochemical modifications push $p_0$ toward $1$, where the Z-channel ceiling $\bm{C}_{\max}(p_0)$ approaches $1$ bit (Table~\ref{tab:cmax_ceiling}). This shows that high channel capacity in this regime is biophysically accessible via receptor-composition modifications, though the wild-type operating points do not sit at these high-capacity parameter combinations. For simplicity, in its channel capacity calculations, this study has ignored the effect of adaptation on performance. The hope was that if indeed the perpetually low concentration regime were relevant \cite{celani2010bacterial}, this calculation would be biologically relevant.

Our results suggest that future work should relax the assumption that adaptation is not relevant. Indeed, one way of viewing adaptation is that it translates a small dynamic range into a much larger range by recentering the response to new baseline concentrations, thereby correspondingly increasing information transfer as has been shown for neural systems \cite{brenner2000adaptive}. A channel capacity calculation for chemotactic sensors with adaptation is far more complicated, and the temporal dynamics of the concentration input will have to be correctly characterized in order to understand the timescale of the input changing versus the timescale of response. Such temporal characterization is a genuinely tough undertaking, but has been done for other biological problems such as with natural scene statistics \cite{olshausen1997sparse,ruderman1993statistics}.

We hope that even in such regimes, the static channel capacity used here is still relevant because it is far easier to estimate from experiments than the trajectory mutual information rate, which faces a curse of dimensionality as trajectory length increases. The bound we establish applies to the activity trajectory as read out on the perceptual timescale $\tau$. With $\tau = 0.22\ \mathrm{s}$, it lies well above information rates measured to date, so it is loose in practice, and where the true rates sit relative to it is an empirical question.


This lends another information-theoretic analysis to the long list of information-theoretic and optimality analyses \cite{mattingly2021escherichia,tjalma2023trade,celani2010bacterial}, although we are far from claiming that an optimality principle explains \textit{E.\ coli}. Other work has shown that efficient prediction as in Ref. \cite{tjalma2023trade} also explains cultured neurons spiking \cite{lamberti2023prediction}, human behavior \cite{ferdinand2024humans}, and salamander retinal ganglion neuron spiking \cite{palmer2015predictive}. It is possible that all these optimality principles are required to explain \textit{E.\ coli}, or that high performance on one optimality principle is correlated with high performance on another. Further tests are required, especially studies of naturalistic chemoattractant concentration statistics seen by bacteria, to understand whether or not information processing capabilities in bacteria explain their performance.


\begin{acknowledgments}
Research was sponsored by the Army Research Office and was accomplished under Grant Number W911NF-25-1-0260. The views and conclusions contained in this document are those of the authors and should not be interpreted as representing the official policies, either expressed or implied, of the Army Research Office or the U.S. Government. The U.S. Government is authorized to reproduce and distribute reprints for Government purposes notwithstanding any copyright notation herein. All code is available at https://github.com/Hail-Earendil/bacterial-chemotaxis-7d-sweep.
\end{acknowledgments}

\bibliographystyle{apsrev4-2}
\bibliography{apssamp}

\end{document}